\documentclass[journal]{IEEEtran}
\usepackage{cite}
\usepackage{amsmath,amssymb,amsfonts}
\usepackage{graphicx}
\usepackage{textcomp}
\usepackage{xcolor}
\def\BibTeX{{\rm B\kern-.05em{\sc i\kern-.025em b}\kern-.08em
    T\kern-.1667em\lower.7ex\hbox{E}\kern-.125emX}}

\usepackage[hyphens]{url}
\usepackage[hidelinks]{hyperref}
\hypersetup{breaklinks=true}

\usepackage{enumitem}
\usepackage{tabularx}
\usepackage{xspace}
\usepackage{multirow}
\usepackage{booktabs} % For formal tables
\usepackage{amsfonts}
\usepackage{makecell}
\usepackage{amsmath}

\usepackage{algorithm}
\usepackage{graphicx}
\usepackage{mathtools}
\usepackage[noend]{algpseudocode}
\usepackage{color}
\usepackage{array}
\usepackage{pgfplots}
\usepackage{commath}
\pgfplotsset{compat=1.17} 

\usepackage{color,soul}
\usepackage{booktabs}

\DeclareMathOperator*{\argmin}{\arg\!\min}

\usepackage[export]{adjustbox}

\usepackage[caption=false]{subfig}
\let\MYorigsubfloat\subfloat
\renewcommand{\subfloat}[2][\relax]{\MYorigsubfloat[]{#2}}

\usepackage{lipsum} % filler text

\usepackage{setspace}
\usepackage{float}

\usepackage{wrapfig}
\usetikzlibrary{positioning}
\usetikzlibrary{patterns}
\newcolumntype{L}[1]{>{\raggedright\let\newline\\\arraybackslash\hspace{0pt}}m{#1}}
\newcolumntype{C}[1]{>{\centering\let\newline\\\arraybackslash\hspace{0pt}}m{#1}}
\newcolumntype{R}[1]{>{\raggedleft\let\newline\\\arraybackslash\hspace{0pt}}m{#1}}

\algnewcommand\algorithmicinput{\textbf{Input:}}
\algnewcommand\INPUT{\item[\algorithmicinput]}
\algnewcommand\algorithmicoutput{\textbf{Output:}}
\algnewcommand\OUTPUT{\item[\algorithmicoutput]}

\DeclareMathAlphabet{\mathpzc}{OT1}{pzc}{m}{it}

\begin{document}

\title{Are Turn-by-Turn Navigation Systems of Regular Vehicles Ready for Edge-Assisted Autonomous Vehicles?}

\author{Syeda Tanjila Atik, Marco Brocanelli,~\IEEEmembership{Member,~IEEE},
and Daniel Grosu,~\IEEEmembership{Senior Member,~IEEE}
        % <-this % stops a space
\thanks{S. T. Atik, M. Brocanelli, and D. Grosu are with the Department of Computer Science, Wayne State University, Detroit, USA. E-mail: $\{$tanjilaatik, brok, dgrosu$\}$@wayne.edu}% <-this % stops a space
%\thanks{Manuscript received September, 2022.}
}

%\markboth{IEEE Transactions on Intelligent Transportation Systems,~Vol. X, No.~X, September~2022}%


\maketitle
\begin{sloppypar}

\begin{abstract}
Future private and public transportation will be dominated by Autonomous Vehicles (AV), which are potentially safer than regular vehicles. However, ensuring good performance for the autonomous features requires fast processing of heavy computational tasks. Providing each AV with powerful enough computing resources is certainly a practical solution but may result in increased AV cost and decreased driving range. An alternative solution being explored in research is to install low-power computing hardware on each AV and offload the heavy tasks to powerful nearby edge servers. In this case, the AV's reaction time depends on how quickly the navigation tasks are completed in the edge server. To reduce task completion latency, the edge servers must be equipped with enough network and computing resources to handle the vehicle demands. However, this demand shows large spatio-temporal variations. Thus, deploying the same amount of resources in different locations may lead to unnecessary resource over-provisioning.

Taking these challenges into consideration, in this paper, we discuss the implications of deploying different amounts of resources in different city areas based on real traffic data to sustain peak versus average demand. Because deploying edge resources to handle the average demand leads to lower deployment costs and better system utilization, we then investigate how peak-hour demand affect the safe travel time of AVs and whether current turn-by-turn navigation apps would still provide the fastest travel route. The insights and findings of this paper will inspire new research that can considerably speed up the deployment of edge-assisted AVs in our society.
\end{abstract}


\section{Introduction}
\label{sec:intro}
\IEEEPARstart{A}{\lowercase{recent}} report~\cite{litman2020autonomous} estimates that by 2045 as much as half of new vehicle sales could be Autonomous Vehicles (AVs). The main advantages of AVs are their ability to provide increased productivity, reduced driver stress, reduced energy consumption, and increased safety~\cite{schildkrautai}. 
Autonomous driving relies heavily on a variety of sensor-generated data (e.g., radar, camera, lidar, ultrasonic sensor) to identify the environment and carry out safe driving operations automatically. For example, each camera frame must be processed on the most appropriate unit  (e.g., CPU, GPU, AI accelerator) to identify features such as lanes and other vehicles directions. The navigation system then uses this information to react by updating the AV's speed and direction~\cite{wang2019auto}. In order to ensure safety and smooth ride, it is thus crucial for the end-to-end response time (i.e., from camera frame acquisition to AV reaction) to be minimized.

% To carry out the safety measures, a series of camera frames need to be processed in milliseconds so that the surrounding environment can be identified properly on time and precise operation instruction can be given accurately to the steering system. Furthermore, the AVs must make critical
% decisions to avoid accidents in real-time and thus require the execution of advanced computer vision, machine learning, and control algorithms. These algorithms are typically computationally intensive and are executed on highly heterogeneous platforms composed of Graphics Processing Units (GPUs), Central Processing Units (CPUs), AI accelerator Application-Specific
% Integrated Circuits (ASICs) and Field-Programmable Gate Arrays (FPGAs) \cite{grigorescu2020survey}. The current AV platforms do
% not scale well with the increasing number of applications that need to be executed \cite{siegel2017survey} and tasks having heavy computational requirement cannot be done by on-board computing units alone. \cite{lin2018architectural}. 

Given that AV technology is not yet deployed at large scale and it is still being studied in both academia and industry, a debate exists between two different deployment strategies that can help mitigate these challenges. The first one is to equip the AVs with a very powerful computing system, which would make them able to run those heavy tasks locally with a low response time. However, this may
lead to an increased production cost of the AVs, which may slow down their rate of sale and widespread adoption. 
% To carry out the safety measures a huge volume of data from the sensors need to be processed with a very low latency which may not be possible to do in-vehicle due to constrained on-board computing capacity \cite{zhuang2019sdn} \cite{zhang2018vehicular}.
The second strategy is to assist the AVs with edge technology, where the AVs can offload some heavy tasks to powerful nearby edge servers~\cite{ben2020edge, cui2020offloading}, and achieve faster data processing with lower response time~\cite{ren2019survey}~\cite{li2020energy}. Compared to the first strategy, the capital cost necessary to install/maintain such servers (e.g., through government incentives) can be amortized by the fact that they are shared across multiple AVs, can help reduce AVs’ cost, and may help speed up the transition to a safer and more efficient traffic circulation. However, this strategy also leads to additional challenges that must be further explored. 

When data is being offloaded to the nearby edge servers, the analysis or processing of the data vastly depends on the amount of computational resources deployed in that edge server. Since the AV keeps moving after offloading its data, it may travel some distance before the computing result from the edge server is received. We call this distance, the \emph{blind distance} since the AV remains \textit{blind} to new features found in the last data offloaded until the computation result is received. With AVs, the blind distance can be bounded \emph{by design} to a certain value (e.g., 1 meter). This requirement can be translated into a deadline for the offloaded task through the AV's speed (i.e., deadline equals blind distance over speed), which can vary over time. As a result, the amount of computing resources deployed at the edge responsible for processing the offloaded task can limit the AV's \emph{safe speed} that allows the vehicle to meet the chosen blind distance requirement.

% as \textbf{Blind Distance}. 
% This blind distance can be bounded to certain levels which also bounds the deadline of each navigation related task to be completed to ensure certain level of safety. For a certain safe blind distance, the AVs must move in a certain speed to provide enough reaction time to unexpected events. We define this speed as \textbf{safe speed}. 

% Furthermore the reaction time depends on how quickly the navigation tasks are completed in the edge server.

In order to reduce the response time and maximize the safe speed, the edge servers must be equipped with enough network and computing resources to handle the vehicle demands. However, because of the variable number of vehicles on the road,  this demand is characterized by large spatio-temporal variations.
As a result, deploying the same amount of resources to all edge servers and/or deploying the resources necessary to handle the peak-hour demand in a specific area may lead to costly and unnecessary over-provisioning of edge resources. On the other hand, deploying a lower amount of resources to limit costs may lead to lowering the safe speed of AVs on a specific path in order to keep a desired bounded blind distance. This may lead to the problem that modern turn-by-turn navigation systems, which provide the fastest route to destination for regular (human-operated) vehicles, may not be able to provide the fastest route for edge-assisted AVs. 

In this paper, we explore the design trade offs among the amount of deployed edge resources, desired blind distance, and resulting safe speed for edge-assisted AVs. First, we design an algorithm that finds the minimum necessary amount of network and computing resources that must be deployed in a specific area to satisfy the hourly demand of a certain number of vehicles. The deadline requirement for each AV is calculated based on a desired input blind distance and average speed at a certain time of the day. Then, we use a real vehicle transit dataset~\cite{uppoor2013generation} from the city of Cologne, Germany, to run our algorithm and find the peak and average configurations of edge resources that satisfy the daily peak and average demands, respectively. We find that, when high safety is required (i.e., short blind distance), the peak configurations lead to high capital costs and high over-provisioning due to the high variability of traffic during the day. Thus, cost savings and better utilization can be achieved by deploying the average configuration. Finally, we study the effect of deploying average configurations on the safe speed of autonomous vehicles for various safety requirements and study its effect on the travel time for random routes in the city. \emph{We find that, due to limitations on available edge resources, modern turn-by-turn navigation systems that provide the fastest route to regular vehicles do not necessarily provide the fastest one for edge-assisted AVs.} We hope the discussion and findings of this paper will inspire and motivate future research on AV navigation systems and algorithms, and help determine how to plan resource deployment for AVs. 

Specifically, this paper makes the following contributions:

\begin{itemize}
    % \item We investigate if the turn by turn navigation systems developed for regular vehicles would work in case of edge assisted AVs.
    \item To the best of our knowledge, this is the first paper to study the design trade offs for edge networks assisting the navigation of AVs and its effects on AV's travel time at different locations and time of the day.
    \item We have used real traffic data from the city of Cologne, Germany to study the traffic characteristics, e.g., number of vehicles and average speed, to find a likely configuration of the resources needed at the edge server located at a particular area to ensure a certain level of safety.
    \item We design an algorithm that explores various design choices and finds the minimum required edge resource configuration in terms of wireless channel capacity and number of logical computing cores to satisfy the peak demand in different areas of the city, i.e., no vehicle exceeds the desired blind distance.
    % \item  We have discussed the implications of deploying wireless and computing resources at the edge to satisfy peak demand vs average demand

    \item We have created several scenarios and compared how the travel time through the same route changes for regular vehicles and edge-assisted AVs. We have shown that in some cases the current turn-by-turn navigation system fails to provide the fastest route for the AVs because they do not consider the amount of resources deployed at the edge servers and the traffic loads at different times of the day in the areas comprising the routes.
 
\end{itemize}

The rest of the paper is organized as follows. Section~\ref{sec:related} reviews the related work. Section~\ref{sec:overview}  provides an overview of edge-assisted navigation for AVs. Section~\ref{sec:process} describes the procedure of finding the configuration of edge resources. Section~\ref{sec:results} describes the dataset along with the results from our experiments. Section~\ref{sec:discuss} discusses the experimental results and Section~\ref{sec:conclude} concludes the paper.

\section{Related work}
\label{sec:related}
Given that AVs are still in their infancy and have not been deployed at large scale yet, recent research studies have envisioned they will likely be designed in one of two ways. First, all the on-board sensor data is exclusively analyzed locally by deploying powerful computing resources on each AV. Second, to reduce the amount of computing resources deployed on each AV, part of the heavy computation is offloaded from AVs to powerful edge resources shared among nearby AVs.

The research presented in~\cite{knuth2009distributed, rekleitis2002multi} consider each vehicle having a full sensor configuration that can navigate on its own without the need for any cooperation with the other vehicles. Research presented in~\cite{asvadi20163d, 6957799, broggi2013full} focus on improving the artificial perception, which is the process of transforming the sensor data into a model to effectively define the surrounding environment. They mostly considered data from a 3D-LiDAR (Light Detection and Ranging) mounted on board of the vehicle. To implement Simultaneous Localization and Mapping (SLAM) in AVs, LiDAR point clouds maps are coupled with camera-images and RADAR (Radio Detection and Ranging ) data~\cite{masmoudi2019object}, \cite{cadena2016past} to assist the control system of the AV to safely navigate through dynamic environments~\cite{masmoudi2019autonomous}. However, the LiDAR sensors are generally expensive and the resulting computation necessary to interpret, maintain, and fuse data in real-time is most likely going to need power-hungry onboard components such as graphics processing units (GPU)~\cite{venugopal2013accelerating}. However, according to a recent study in~\cite{lin2018architectural}, adding too many resources on board can have consequences on the driving range of the vehicle.

In order to deal with these challenges, several studies suggest computational offloading. Several researchers~\cite{wright2020cloudslam, ben2020edge, ashok2016adaptive, cui2020offloading}, propose to offload part of the heavy computation to the edge. Some studies~\cite{lucic2020latency, jayaweera2019autonomous, lucic2019generalized, 8885276} even proposed to move the LiDARs out of the vehicle and place them in other elevated positions such as side of the buildings or lamp posts and move the GPUs at the edge by connecting them with edge servers so that they can be shared among other AVs. However, it remains largely unclear what are the consequences of edge-resource deployment choices on AVs. \textit{To the best of our knowledge, this is the first work studying the design trade offs of edge resources for future AVs and their consequences in terms of safety and travel time compared to regular vehicles.} We hope the discussion and findings of this paper will inspire new research directions on edge-assisted AVs.

\section{Edge-Assisted Navigation}
\label{sec:overview}
\begin{figure}[!t]
\vspace{0.1in}
\centering
\includegraphics[width=0.9\columnwidth]{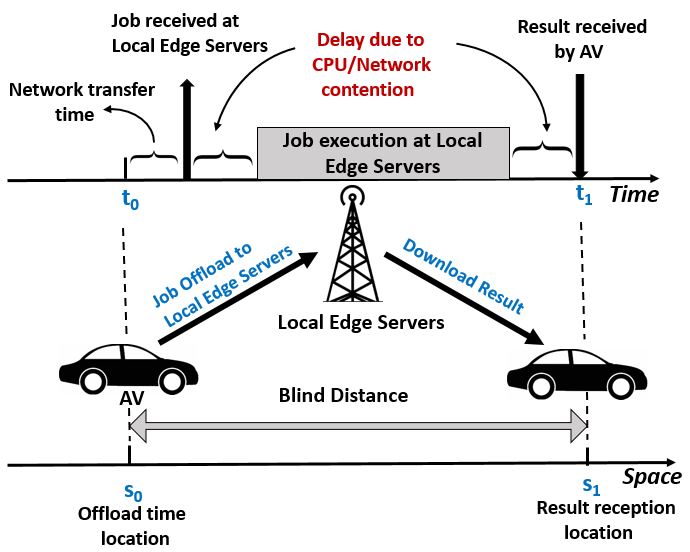}
\caption{\label{fig:system} Overview of the considered system scenario}
\vspace{-0.15in}
\end{figure}

In order to conduct our qualitative analysis, we consider a system scenario where an edge server is placed at the center of an area. The network and computing resources of that server are shared among all the AVs in that area. Figure~\ref{fig:system} shows our example system scenario. An AV offloads its job (computation) to the nearby edge server while being in location~$s_0$ at time~$t_0$ by using the underlying communication network. After a certain transfer time~$t$, the job is received at the edge server where it is scheduled concurrently with other AVs' jobs, based on a certain policy. After it finishes its execution, the result is returned to the AV at time~$t_1$ in location~$s_1$ and the AV reacts according to the received result, e.g., brake or steer. The distance that the AV travels between job offloading and result reception is defined as the \emph{blind distance},~$L$. The time taken for a job to finish processing at the edge server from the time it is offloaded is defined as the \emph{response time},~$r$ of that job. When the AV receives the result of the previously offloaded job it acquires a new sensor data and offloads another job. Note that in this paper we want to determine the conditions to satisfy a certain blind distance requirement, so considering the AVs to offload jobs periodically rather than sporadically would not change the results of our study.

\subsection{Assumptions} In order to simplify the problem for the ease of~understanding, we have made the following assumptions:
\begin{itemize}
    % \item All the AVs  have similar/uniform navigation system component and   the computation related to that component relies on the edge server.  In real life we can have heterogeneous job offloading but in this paper we want to have a qualitative   analysis of what would be the case if only navigation related computation is   
    % offloaded. Having heterogeneous jobs and data sizes are not going to invalidate the results of our analysis as the end result would have similar trends.
    
    \item All the AVs  have the same autonomous navigation software and the related computation relies on the nearby edge servers. Thus, we assume that, by design, all AVs offload the same amount of per-job computation. Given the complexity of developing a reliable AV navigation software, car manufacturers such as General Motors (GM) are likely going to use the same autonomous navigation software across different vehicles to improve reliability and lower the development complexity. We also extend this assumption across car manufacturers. On the other hand, having heterogeneous jobs offloaded to the edge server is unlikely to change the trend of the results presented in this paper since (1) the algorithm developed to find peak and average configurations would adapt network and computing resources accordingly, and (2) the main reason for the high variations in edge utilization for peak configuration is due to large hourly traffic variations during each day, which is likely to be consistent with regular or autonomous vehicles.

    % On the other hand, there could be heterogeneous job offloading. In this case the amount of edge resources required can be different but the behavior will be the same as we have found from our experiments. Our key point of consideration is efficient provisioning of edge resources with lower capital cost.

    % but in this paper we want to have a qualitative   analysis of what would be the case if only navigation related computation is   
    % % offloaded. Having heterogeneous jobs and data sizes are not going to invalidate the results of our analysis as the end result would have similar trends.

    \item Each edge server is equipped with a number of generic logical cores and each of them are able to carryout the requested computation within a bounded worst case execution time. We make this assumption to abstract the complexities of dealing with specific CPU types, frequency, and other hardware characteristics. Each logical core could be translated to a specific CPU type with a certain frequency by the system designer.
    % \item The AVs are connected with the RSU located in the area they are residing in through a wireless network interface. The total network bandwidth is equally shared by the number of AVs connected to the RSU at a particular time in that particular area while jobs are being offloaded.
    \item The AVs are connected through a wireless network
    interface with the edge server located in
    the area they are residing in. To keep our study general, we determine the required total channel capacity, which is the maximum information rate that a channel can transmit~\cite{channel}. This will be equally shared by the number of AVs connected to the edge server at a particular time in that particular area while jobs are being offloaded. The channel capacity can be used to derive the amount of required bandwidth considering  signal and noise power. Those parameters depend on the particular network interface the system designer decides to use. 
    \item We assume that each edge server schedules jobs according to the non-preemptive Earliest Deadline First (EDF) scheduling policy~\cite{mok1978multiprocessor}, which is typically used in real-time systems. Specifically, our system tasks are modeled as a constrained-deadline sporadic task model. We choose this policy because it is commonly used to study the behavior of real-time systems. Other schedulers could easily be employed for testing.
    \item Similar to related work~\cite{8712145} and~\cite{chen2021enhancing}, due to the data size of the result of the computation being very small, we assume that the time to send back the results to the AV is negligible.
    \item As we do not have the speed limit of the roads in the dataset, we assume that the regular vehicles travel at the maximum speed allowed when the roads are free and at the traffic speed when there is traffic. For that reason, we assume, the AVs cannot exceed the average speed of regular vehicles recorded in the dataset at each specific hour and area, which avoids exceeding speed limits or the real traffic speed.
\end{itemize}

%%example system scenario figure

% -We have assumed that only the computations that cannot be executed locally within the required deadline will be offloaded to the edge and the result will be sent back to the requesting vehicle when completed. server.\newline
% - each vehicle offloads a job to the edge server as soon as it receives the result of the previously offloaded job.\newline
% -Each edge server is equipped with a number of generic logical cores each of them are able to carryout the requested computation at a bounded worst case execution time. Each logical core could be translated to a specific CPU type with a certain frequency.\newline
% -The AVs are connected with the edge server or RSU located in the area they are residing in through a network. The total network bandwidth will be equally shared by the number of AVs connected to the RSU at a particular time in that particular area while data are being offloaded.\newline
% -As the data size of the result of the computation is very small we assume the time to send back the results to the AV to be negligible. Research works related to computation offloading in \cite{8712145} \cite{chen2021enhancing} assume the same.

Based on the above assumptions and setup, we have first analyzed the hourly average traffic speed and vehicle count from a comprehensive real dataset. Then, we varied the required blind distance to obtain a specific deadline for each city area and hour. Based on that deadline, we compute the total network channel capacity and the number of logical cores required for the completion of all the jobs of the AVs within the deadline, by using Algorithm~\ref{alg:config}. We call this the \emph{required
configuration} of resources and calculate it for every hour of the day for the edge server residing in a particular area. The entire process is described in the next section. %Section~\ref{sec:process}.

\section{Configuration Search}
\label{sec:process}
In this section, we describe a configuration search algorithm that  determines the minimum amount of edge network and computing resources necessary to satisfy the total vehicle demand for a fixed maximum blind distance in a certain geographical area at a specific time of the day. The notation used in the paper is summarized in Table~\ref{tab:notation}.

%%table goes here%%%%%%%%%
\begin{small}
\begin{table}[t!]
\vspace{-0.1in}
	\centering
	\caption{Notation}\vspace{-0.1in}
	\label{tab:notation}\
	\begin{tabular}{ll}
		\hline
		Notation & Description \\  \hline
		$L$ & blind distance\\ 
		$d$ & relative deadline of a job\\
		${\delta}_i$ & absolute deadline of a job of vehicle $i$\\
% 		${t}$ & particular time of the day\\ 
		$V$ & average number of vehicles \\
		$S$ & average speed of the vehicles\\ 
		$D$ & data size \\ 
% 		$\mathcal{B}$ & Network bandwidth\\
		$E$ & worst-case job execution time at edge \\ 
		$b$ & required total channel capacity\\
		$c$ & number of logical cores \\

		$\Delta^b$ & increment in the channel capacity\\
		$\Delta^c$ & increment in the number of logical cores\\
		
		${d^m}$ & number of deadline missed jobs\\
		$r$ & response time of a job \\
		$r^{max}$ & maximum response time of a job \\
	
		$\Delta^{r}$ & response time variation \\
		$\epsilon$ & response time variation threshold\\
		$r^{temp}$ & temporary response time of a job\\
		
		${t^{max}}$ & maximum allowable transfer time\\
% 		${T_r}$ & transfer rate\\
		${t}$ & current transfer time\\
		$W$ & duration of the working period\\
		${k}$ & current time unit of the schedule\\
	    ${V^{c}}$ & number of vehicles per logical core\\
		${o_i}$ & time when a job  of vehicle $i$ gets offloaded\\
		${a_i}$ & job arrival time for vehicle $i$\\
		${s_i}$ & job processing start time for vehicle $i$\\
		${f_i}$ & job processing completion time for vehicle $i$\\
\hline
	\end{tabular}%\vspace{-0.2in}
	
\end{table}
\end{small}

%%%%%%%%%% algoritm 1
\alglanguage{pseudocode}
\begin{algorithm}[t!]
%\DontPrintSemicolon
{
    \begin{algorithmic}[1]
    
    \INPUT{$L$, blind distance\newline
    \hspace*{\algorithmicindent}$V$, average number of vehicles\newline
    \hspace*{\algorithmicindent}$S$, average speed of the vehicles\newline
    \hspace*{\algorithmicindent}$D$, data size\newline
    \hspace*{\algorithmicindent}$E$, processing time at the edge\newline
    \hspace*{\algorithmicindent}$W$, total working period\newline
    \hspace*{\algorithmicindent}$\eta$, multiplicative factor to account for queuing delay 
    \hspace*{\algorithmicindent}\hspace*{0.5cm}at edge server\newline
    \hspace*{\algorithmicindent}$\Delta^b$, increment in the channel capacity\newline
    \hspace*{\algorithmicindent}$\Delta^c$, increment in the number of logical cores\newline
    \hspace*{\algorithmicindent}$\epsilon$, response time variation threshold\newline
    \hspace*{\algorithmicindent}$M, N$, sufficiently large numbers}
    % \OUTPUT{$\mathcal{A}, \mathcal{C}$}
   
    \State{$d^{m} \gets M $}\Comment{Number of deadline misses}
    \State{$d \gets L/S  $}\Comment{Relative deadline}
    \State{$t^{max} \gets  d- \eta E$}\Comment{Maximum allowable transfer time}
    \State{$b \gets (D \cdot {V})/ t^{max}  $}\Comment{Total channel capacity}
    % \State{$a \gets a - \Delta_a $}
    % \State $offset \gets 0$
    \While{$d^{m} > 0$}
        % \State{$a \gets a + \Delta_a $}
        % \State {$t^{rate} \gets b/{V_t}$}
        \State{$t \gets (D \cdot V)/ b $}\Comment{Current transfer time}
        % \State{$c \gets 1 - \Delta_c + offset  $}
        \State{$c \gets 1 $}\Comment{Number of logical cores}
        % \If{$c < 0$}
        %     \State{$c \gets 1$}
        % \EndIf    
        \State{$r^{max} \gets 0$}\Comment{Maximum response time}
        \State{$\Delta^{r} \gets N$}\Comment{Response time variation}
        \While{$d^{m} > 0$ \textbf{and} $\Delta^{r} > \epsilon $}
            \State{$d^{m}, r^{temp} \gets \textsf{SCHED}( c, t, E, V, d,  W) $}
            \State{$ \Delta^{r} \gets \abs{(r^{max} - r^{temp})/r^{temp}} $}
            \State{$r^{max} \gets r^{temp} $}
            \If{$d^{m} > 0$ \textbf{and} $\Delta^{r} > \epsilon$}
                \State{$c \gets c + \Delta^c$}
            \EndIf    
        \EndWhile
        
        % \State{$offset \gets c - \Delta_{off} $}
        \If{$d^{m} > 0$}
            \State{$b \gets b + \Delta^{b}$}
        \EndIf    
        
    \EndWhile
    % \State{$\mathcal{A} \gets a$} 
    % \State{$\mathcal{C} \gets c$} 
    \OUTPUT{$b, c$}

	\end{algorithmic}
}	

\caption{\textsf{Configuration-Search}}
\label{alg:config}
\end{algorithm}

%%%%%description of algorithm 1
\subsection{Finding configuration of resources}
Algorithm \ref{alg:config}, \textsf{Configuration-Search} describes the procedure of finding the required amount of edge resources for a particular area, time of the day, and blind distance.

The algorithm first initializes the number of deadline misses to a sufficiently large number~$M$, which lets the while loop in Line~5 to execute at least the first iteration. 
In Line~2, the deadline for every job offloaded by the AVs is calculated based on the input blind distance~$L$ and input average speed of the vehicles~$S$. In Line~3, the maximum allowable time~$t^{max}$ for a job to get transferred to the edge server after it is offloaded is calculated. Here, the parameter~$\eta \ge 1$ accounts for any queuing delay that may occur at the edge server before the job starts executing, e.g., due to contention with jobs of other AVs. In Line~4, the initial total channel capacity~$b$ required for satisfying the maximum transfer time~$t^{max}$ for all vehicles served by the same local edge servers is calculated. The algorithm is searching for the minimum configuration necessary to have zero deadline misses, so it starts by evaluating the system with an initial total channel capacity~$b$, which is used to calculate the current transfer time~$t$, in Line~ 6, and one logical core, in Line~7. 

% The While loop in Line~5-16 gradually increases the total required bandwidth $b$ and the number of logical cores $c$ until we get 0 deadline misses. 
In Line~8, the maximum response time~$r^{max}$ is initialized to~0. In Line~9, the response time variation~$\Delta^r$ is initialized to a sufficiently large number~$N$, which lets the while loop in Line~10 to execute the first iteration. The algorithm then uses the \textsf{SCHED} algorithm to find the maximum response time~$r^{max}$ and the number of deadline misses~$d^m$, by calling it with the current value of transfer time~$t$, current number of logical cores~$c$,  the average number of vehicles~$V$, and the deadline calculated in Line~2. The vehicle jobs that are offloaded are processed at the edge server according to the \textsf{SCHED} algorithm, which is discussed in the next subsection. The while loop in Lines~10-15  keeps increasing the number of logical cores by a factor of~$\Delta^{c}$ (in Line~15) in each iteration until it leads to zero deadline misses or it finds that increasing the number of logical cores~$c$ does not improve the performance anymore. To check for the second condition, the response time variation~$\Delta^r$ is calculated in Line~12 to find how the response time changes compared to the previous iteration. If the response time variation~$\Delta^r$ is less than a threshold~$\epsilon$, we need to increase the total channel capacity~$b$ before we increase the number of logical cores~$c$ again. Hence, the total channel capacity~$b$ is incremented by a factor of $\Delta^{b}$ in Line~17 and the first while loop in Line~5 starts executing again with the new total channel capacity~$b$ and the number of logical cores~$c$ is increased as previously described. This process (Lines~6-17) is repeated and the network and computing resources are increased in each iteration until it obtains zero deadline misses. When it terminates, the algorithm returns the amount of total channel capacity~$b$ and the number of logical cores~$c$ needed to ensure that all the vehicle’s offloaded jobs are processed within the deadline. 

\alglanguage{pseudocode}
\begin{algorithm}[t!]
{
    \begin{algorithmic}[1]

    \INPUT{$c$, number of logical cores\newline
    \hspace*{\algorithmicindent}$t$, transfer time\newline
    \hspace*{\algorithmicindent}$E$, processing time at the edge\newline
    \hspace*{\algorithmicindent}$V$, average number of vehicles\newline
    \hspace*{\algorithmicindent}$d$, relative deadline of the jobs\newline
    \hspace*{\algorithmicindent}$W$, total working period}
    % \OUTPUT{${d_m}, r$}
    \State{${k} \gets 0  $}\Comment{Current time unit of the schedule}
    \State{${d^{m}} \gets 0  $}\Comment{Number of Deadline misses}
    \State{$r^{max}\gets 0$}\Comment{Maximum response time}
    \State{${V^{c}} \gets \lceil{V}/c\rceil $}\Comment{Number of vehicles per logical core}
    \State{Let $v_i$ indicate a specific vehicle $i$.}
   
    %\State{create a queue of ${V_c}$ vehicles}
    \For{$i \in [1,V^c]$}
        \State{${o_i} \gets 0$}\Comment{Time of first offload of vehicle $i$ }
    \EndFor
    \While{${k} < W$}
        \For{$i\in[1,V^c]$}
            \State{${a_i} \gets t + {o_i}$}\Comment{Job arrival time at the edge}
            \State{${\delta}_i \gets {o_i} + d$}\Comment{Absolute deadline of the job}
        \EndFor
        \State{$S\gets\{v_i|a_i\le k, i\in[1,V^c]\}$}
        \If{$S=\emptyset$}
            \State{$k\gets k+1$}
        \Else
            \State{$j\gets \argmin_{v_i\in S}\{d_i\}$}\Comment{Earliest deadline Job}
            \State{${s_j} \gets k$}\Comment{Job processing start time of vehicle $j$}
            \If{${s_j} > W -E$}
                \State{\textbf{break}}
            \EndIf
            \State{${f_j} \gets {s_j} + E$}\Comment{Job completion time of vehicle $j$}
            \State{$k \gets {f_j} $}
            \If{${f_j} > {\delta}_i $}
                \State{${d^{m}} \gets {d^{m}} +1$}
            \EndIf       
            \State{$r \gets {f_j} - {o_j}$}
            \State{$r^{max}\gets \max\{r,r^{max}\}$}
            \State{${o_j} \gets {f_j}$}
            % \State{${Curr} \gets {f_j} $}
        \EndIf
        % \If{$flag = = 1 $}
        %     \State{\textsf{Break}}
        % \EndIf
        % \For{$i \in {V}$}
        %     \State{$\mathpzc{d_i} \gets {o_i} + d$}
        %     \State{${s_i} \gets \textsf{max}({Curr}, {a_i})$}
        %     \If{${s_i} > W -E$}
        %         \State{$flag \gets 1  $}
        %         \State{\textsf{Break}}
        %     \EndIf
        %     \State{${f_i} \gets {s_i} + E$}
        %     \State{${Curr} \gets {f_i} $}
        %     \If{${f_i} > \mathpzc{d_i} $}
        %         \State{${d^{m}} \gets {d^{m}} +1$}
        %     \EndIf    
        %     \State{$res \gets {f_i} - {o_i}$}
        %     \State{${o_i} \gets {f_i}$}
        % \EndFor   
    \EndWhile
    \OUTPUT{${d^{m}}, r^{max}$}

	\end{algorithmic}
}	

\caption{\textsf{SCHED}}
\label{alg:sche}
\end{algorithm}

%%%description of algorithm2 
\subsection{Processing of jobs at the local edge servers}
 Algorithm~\ref{alg:sche}, \textsf{SCHED}, describes the process of how offloaded jobs from each AV are processed at the local edge servers. It selects the vehicle job that arrives at the edge server according to EDF scheduling policy and returns the number of deadline misses and maximum response time of the vehicle jobs at the end of the working period. It first initializes the current time unit of the schedule~$k$, number of deadline misses~$d^m$, and maximum response time of a job~$r^{max}$ to~0 (Lines~1-3). In Line~4, based on the input average number of vehicles~$V$ and the required number of logical cores~$c$, the number of vehicles~$V^{c}$ served by each logical core is calculated. For simplicity, we consider the job processing and job response time at the heavily loaded logical core. That is the logical core that serves the highest number of vehicles assuming a balanced vehicles-per-core allocation, which is why we use the ceiling function in Line~4. In Line~7, the algorithm initializes the  time~$o_i$ when the jobs are offloaded by the vehicles~$V^c$ to~0. In Line~10, the arrival time of the vehicle jobs at the edge server queue,~$a_{i}$ is calculated based on the job offload time,~$o_i$ and job transfer time to the edge,~$t$. In Line~11, the absolute deadline of the jobs~${\delta}_i$ is calculated based on the job offload time~$o_i$ and the relative deadline~$d$. Here the subscript~$i$ denotes a particular vehicle~$v_i \in V^c$.
 
 In Line~12, the vehicle set~$S$ is initialized with the vehicles whose job's arrival time~$a_i$ is within the current time slot~$k$. Among the vehicles in the set~$S$ the vehicle~$j$ having a job with the earliest deadline is chosen to be processed and the current time slot~$k$ is updated to when the job processing is finished. As long as the vehicle set~$S$ is not empty, the jobs of the vehicles in this set are processed in this manner. This is done in Line~16 to Line~21. If the job completion time~$f_i$ is greater than its absolute deadline~${\delta}_i$, the count for deadline missed jobs~$d^{m}$ is incremented by 1 in Line~23. Every time a job is finished processing, its response time~$r$ is calculated in Line~24. Among all the response times, we consider the maximum response time~$r^{max}$ (Line~25). The next job offload time~$o_j$ of vehicle~$j$ is calculated when its previous job finishes its processing at the edge server (Line~26). When the vehicle set~$S$ becomes empty, the current time slot is advanced by 1 unit (Lines~13-14). The process is repeated until the end of working period (the while loop in Line~8 ensures that). Finally, the algorithm returns the maximum job response time~$r^{max}$ and the total number of jobs that missed deadlines,~$d^{m}$.

 \subsection{Determining the performance of edge-assisted AVs}
 From the required edge resource configuration of each hour of the day (found using Algorithm~\ref{alg:config}), we have calculated two types of configurations. The \emph{peak configuration} is calculated as the maximum required total channel capacity and number of logical cores over all the hours. That means the peak configuration handles the peak demand without causing any deadline misses of the vehicle jobs.
The \emph{average configuration} is calculated as the average of the required total channel capacity and the number of logical cores over all the hours. The average configuration handles the average demand.
Afterwards, in order to determine the edge assisted AV’s performance with the average configuration, we created a simulation that uses Algorithm~\ref{alg:sche} to schedule jobs at the edge server. We have used the transfer time~$t$ calculated from the total channel capacity and number of logical cores~$c$ found in the average configuration to obtain the maximum response time~$r^{max}$ of the jobs at each hour of the day for certain blind distances,~$L$. Then, we calculate the AV's safe speed  at a particular time considering a particular blind distance~$L$ as follows
$${safe\_speed} = L/r^{max}$$
Note that, the average configuration handles the average demand, but it may lead to deadline misses if AVs travel at regular vehicle speed, especially at rush hours. Thus, for every deadline miss we use the maximum response time to calculate what the maximum safe speed of the AV should have been to meet the required blind distance. Furthermore, we test the effect of various blind distances on the requirement of network and computing resources, and on the AV's safe speed in terms of those two configurations. We have used this safe speed to further investigate the travel time of the AVs in several routing scenarios. 

\section{Implications of Edge-Assisted AVs}
\label{sec:results}
In this section, we leverage the algorithms described in Section~\ref{sec:process} to conduct a qualitative study on the implications of assisting AVs from the edge. Specifically, we first describe the real-world dataset used to conduct our study and evaluate the hourly number of vehicles and their average speed in a real city. Second, we use the data from this dataset to execute Algorithm~\ref{alg:config} and compare the peak and average resource configurations in terms of total channel capacity and number of logical cores to deploy in different areas of the city. Third, we study how using the average configurations affects the AVs' response time and evaluate the safe speed to satisfy a desired blind distance. Finally, we examine the effect of the safe speed on the travel time of edge-assisted AVs compared to that of regular vehicles using the navigation routes provided by modern turn-by-turn navigation apps. 

\subsection{Dataset description} 
\begin{figure}[!t]
%\vspace{0.1in}
\centering
\includegraphics[width=0.8\columnwidth,valign=c]{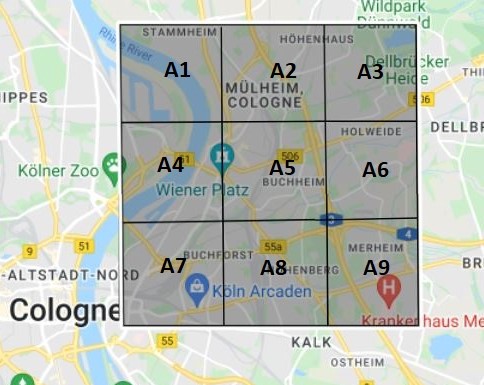}
\caption{The selected nine area locations in Google maps}
\label{fig:areas}
\vspace{-0.15in}
\end{figure}

In order to ensure that our experiments are based on realistic data, we have used the vehicular mobility dataset of Cologne, Germany~\cite{uppoor2013generation}. The dataset includes 24 hours of synthetic car traffic traces comprising an area of 400 square kilometers of a typical working day. Although the dataset is from 2013, it is quite complete because it contains timestamps, anonymized vehicle IDs, vehicle speed, longitude and latitude coordinates of vehicles with a one second time granularity. We have decided to use this dataset because, unlike other newer datasets, it is not restricted to a certain kind of vehicle such as bus or taxi, and includes data of vehicles traveling through both minor and major city roads. Due to the vast amount of data in the dataset, we extracted data for nine sample areas  including the heavily congested ones. We call those areas A1 through A9 as shown in Figure~\ref{fig:areas}. Based on the range of vehicular wireless interfaces (e.g., DSRC), we choose each area to be of 2~km by 2~km in size. We also assume that the edge servers are located in the center of each area.

%%% table of parameter values goes here
\begin{table}[!t]
\small
\caption{PARAMETER VALUES USED IN THE EXPERIMENT}
\label{tab:para_value}
\begin{tabular}{ ll } 
\toprule
 \textbf{Parameter} & \textbf{Value} \\ 
%  \cline{1-2}
\midrule
 Blind distance ($L$) & [2,4,6,8,10,12,16,20] meters \\ 
 Data size ($D^{size}$) & 1.8 Mb\\
 
 Processing time at the edge ($E$) & 16 ms \\ 

 Channel capacity increment ($\Delta^{b}$) & 2 Mbps \\ 

 Logical core increment ($\Delta^{c}$) & 5 \\ 

 Working period duration ($W$) & 60 seconds \\ 

 Initial value of $d^{m}$ and $\Delta^{res}$ & 100\\

$\eta$ & 2 \\

$\epsilon$ & 0.005 \\
\bottomrule

% \label{tab:para_value}
\end{tabular}

\end{table}

%%%%%%%%%

\begin{figure*}[!ht]
% \vspace{-0.15in}
\centering
\subfloat[]{\includegraphics[width=0.45\textwidth,valign=c]{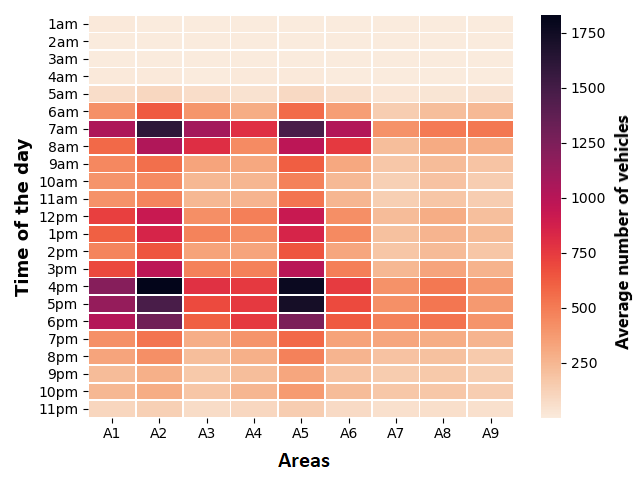}

\label{fig:no of vehicles}

}
\subfloat[ ]{\includegraphics[width=0.45\textwidth,valign=c]{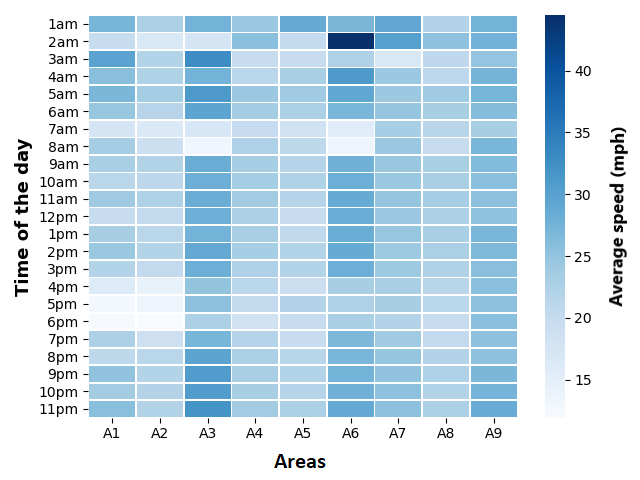}%
\label{fig:speed}
}
%\vspace{0.1in}
\caption{Heatmap of (a) the average number of vehicles and (b) the average speed of the vehicles in the nine areas during the entire day.}
\label{fig:heatmap}
%\vspace{-0.17in}
\end{figure*}

The data we are most interested in studying are the number of vehicles and their average speed for each time of the day. In order to get a snapshot of such variables in each area, we examined the dataset, calculated the average speed of the vehicles, and found that each vehicle would take about three minutes to travel through an area. Hence, we have extracted the number of unique vehicle IDs in the dataset in each three minute period over the first 15 minutes of each hour. Then, we have calculated the average speed and the average number of unique vehicles over the five periods for each area and hour of the day. Figure~\ref{fig:heatmap} shows the heatmaps of the average number of vehicles and average speed for each hour in the nine areas extracted. According to the results in Figure~\ref{fig:no of vehicles}, similar to most major cities, 7~am and 4~pm are the peak rush hours of the day with the highest vehicle count in each area. For example, at 4~pm areas A2 and A5 have an average of 1800 vehicles, areas A3 and A6 have 750 vehicles, and areas A7, A8, and A9 have nearly 400 vehicles. Because of the variable traffic  at different times of the day and areas, the average vehicle speed is also variable. According to Figure~\ref{fig:speed}, during the rush hours the average speed in the area A2 and A5 is less than 16~mph, while areas A3, A6, and A9 have higher speed of 24~mph because of their lower number of vehicles.  In order to analyze the effect of having edge-assisted autonomous vehicles of variable traffic and speed, we have used the data of Figures~\ref{fig:no of vehicles} and~\ref{fig:speed} as inputs for Algorithm~\ref{alg:config}. Specifically, while we examine the general results for all nine areas, in the next sections we are going to provide in-depth results by focusing on areas A3, A5, and A7 since they are representative of moderate, heavy, and low traffic areas, respectively. The values of the parameters used in our experiments are given in Table~\ref{tab:para_value}.

%%fig for bw and logical cores
\begin{figure*}[t!]
% \vspace{-0.15in}
\centering
\subfloat[]{\includegraphics[width=0.45\textwidth]{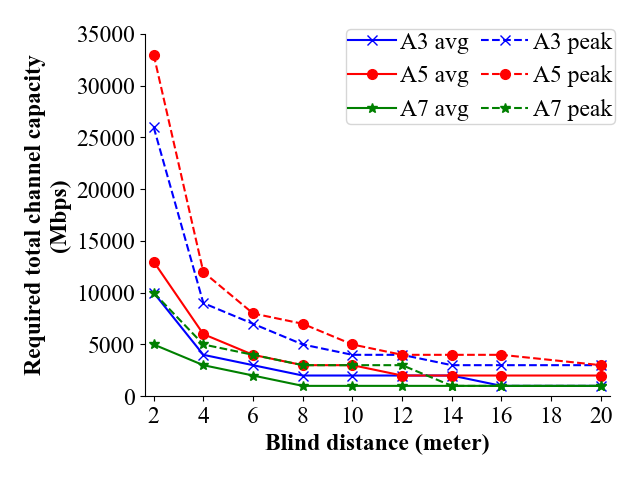}
\label{fig:total bw}
}\hfill
\subfloat[]{\includegraphics[width=0.45\textwidth]{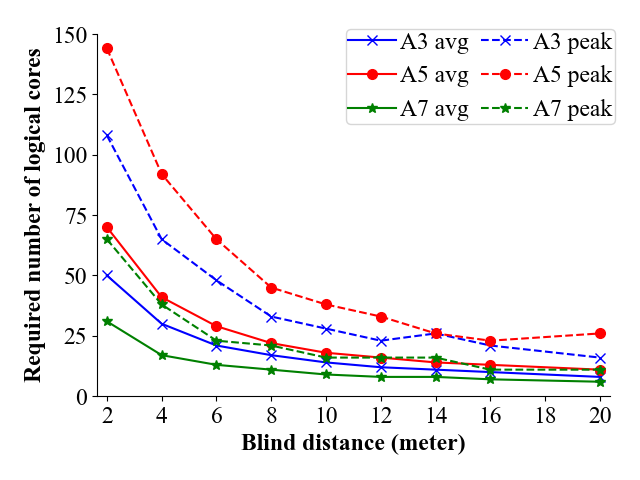}%
\label{fig:cores}
}
%\vspace{0.1in}
\caption{Comparison of (a) the required channel capacity and (b) the required number of logical cores between peak and average configurations for area A3, A5, and A7.}
\label{fig:configuration}
\vspace{-0.17in}
\end{figure*}
%%end of figure

\vspace*{0.1cm}
\noindent\emph{Robustness to dataset.}
We have used the number of regular vehicle data from the above-described dataset in our experiments as it reports the traffic density of different areas at different time. In the near future, when  AVs will be deployed at large scale, we can expect to have similar traffic density with similar traffic patterns, e.g., rush hours in early morning and afternoon.
The algorithms we have described are not tied to any particular values from the dataset. Those can be used for any number of vehicles and we expect that the trend of the results would not change much as there would always be peak hours and traffic congestion on the roads.

\subsection{Edge resource deployment}
In this section, we use Algorithm~\ref{alg:config} to study how the resource deployed at the edge changes for different input requirements and city areas.

%%%need to know how to describe the speed figure.

%fig for stem and response time
\begin{figure*}[!t]
% \vspace{-0.15in}
\centering
\subfloat[]{\includegraphics[width=0.47\textwidth]{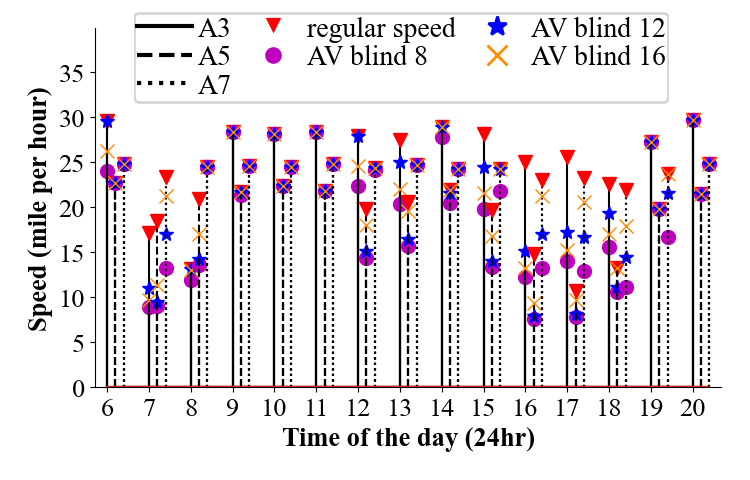}%
\label{fig:stem}
}\hspace*{0.6cm}
\subfloat[]{\includegraphics[width=0.41\textwidth]{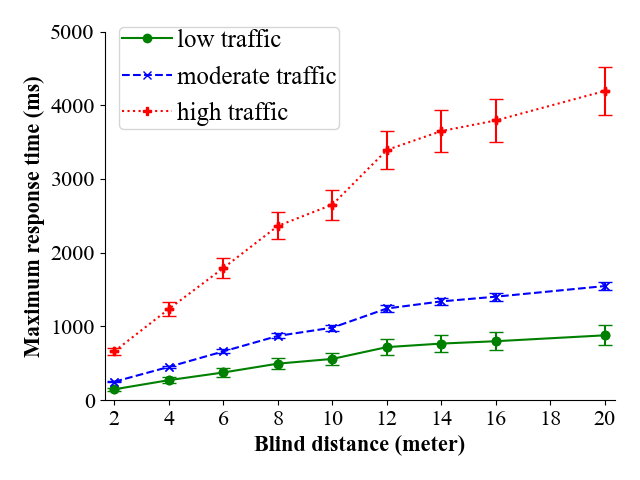}
\label{fig:response}
}
%\vspace{0.1in}
\caption{(a) Comparison of the regular vehicles' speed with the AVs' \textit{safe speed} in the three areas during several hours of the day; (b) Maximum response time for different blind distances with average configuration in area~A5.}
\label{fig:max_res_stem}
\vspace{-0.17in}
\end{figure*}

Figure~\ref{fig:configuration} shows the total channel capacity and the number of logical cores required for different blind distances while considering the average and peak configurations for the three sample areas. A5, A3, and A7, which are representative of higher to lower traffic density, respectively.  
As discussed earlier, the peak configuration can handle the peak demand, in other words, all the demands at any time of the day are handled without any deadline misses. As shown in Figure~\ref{fig:total bw}, 
in the case of peak configuration, the total channel capacity requirement is 33000 Mbps (Mega Bit per Second), 26000~Mbps, and 10000 Mbps for areas A5, A3, and A7, respectively, considering a blind distance of~2 meters. That means area~A5, which is the most heavily congested, requires 27\% and 230\% more channel capacity than areas A3
and A7, respectively. Thus, deploying the same amount of network resources in all city areas would not be a reasonable choice. 
The peak network channel capacity could be considerably reduced by allowing higher blind distances at the cost of a lower AV safety. For example, increasing the blind distance from 2~meters to 10 meters in A5 would reduce the needed peak channel capacity by 85\%. 
Furthermore, considering the average configuration, 60\% and 33\%, a lower channel capacity is required compared to the peak configuration  considering the lowest (2 meters) and  the highest blind distance (20 meters), respectively, for area~A5. Although using the average configuration may lead to slowdown of AVs compared to the regular vehicles during the rush hours to maintain the required blind distance, greater cost savings can be achieved while deploying the network resources compared to the peak configuration. The amount of actual cost savings depends on the network type. For areas A3 and A7, a similar trend can be seen in terms of using the average configuration over the peak configuration.

Figure~\ref{fig:cores} shows the  number of logical cores required for different blind distances considering the two different configurations in those three sample areas. Because of the variable traffic density in areas A3, A5, and A7, using the peak configuration would require 108, 144, and 65 logical cores, respectively, which are enough to sustain the traffic demand at all times of the day. Similar to the network requirements, we observe a large variability in the number of logical cores to be deployed in peak configurations. This further demonstrates the unnecessarily high capital expenditures to deploy a uniform amount of resources city-wide.
Furthermore, considering the peak configuration, 144 logical cores are required for blind distance of 2~meters for area~A5. For the same previously described reason, the number of required logical cores decreases with the increase of the blind distance. However, using the average configuration, the required number of logical cores is much lower, approximately 50\% less than what is required for the peak configuration. Similarly, for a blind distance of 12~meters in area~A5, we would require 33 and 16 logical cores considering peak and average configuration, respectively.

After analyzing the resource requirement for the two configurations, we can say that if we use the peak configuration, it would be beneficial only during the rush traffic hours (specifically 7~am, 8~am, 12~pm, 3~pm, 4~pm, 5~pm, and 6~pm), which represent less than one third of the day. During the other normal hours when the number of vehicles is lower, half of the resources would remain underutilized. On the other hand, if we use the average configuration, although we would get AV slowdown during those rush hours, we can meet the demand during normal hours. This can result in a much better utilization of the resources and help achieving an efficient resource provisioning at the edge.

\subsection{Edge-assisted AVs' performance}
Given the potentially high capital expenditure reductions  of deploying the average configurations in different areas of a smart city, here, we study what would be its effect on the AV's responsiveness and safe speed. Figure~\ref{fig:stem} shows the comparison of the AV's safe speed at blind distance 8~meters, 12~meters, and 16~meters with the average speed of the regular vehicles in areas A3, A5, and A7 for each hour from 6~am to 8~pm. Here, the different lines denote different area and the markers on each line denote the average speed of the regular vehicles and safe speed of the AVs in that particular area considering those three blind distances.

It can be seen that during the medium and low traffic hours such as 10~am, 11~am, 12~pm, and 8~pm, the traffic demand can be met well with the average configuration of resources at the edge server. As a result, there is no slow down of the AV's safe speed compared to the speed of regular vehicles. However, during the rush hours traffic (specifically 7~am, 8~am, 12~pm, 3~pm, 4~pm, 5~pm, and 6~pm), there is some slowdown in the safe speed of the AVs because the average configuration is not able to handle the peak traffic demand. Usually, increasing the blind distance increases the deadline of the vehicle job, which lets the job to have larger response time and still meet the deadline. Figure~\ref{fig:response} shows how the maximum response time increases with increased blind distance for low, medium, and high traffic. We have used the K-Means algorithm \cite{k:2013} to cluster the hours into low, medium, and high traffic hours according to the number of vehicles present during those hours. We have shown this result only for area~A5 as the other areas exhibit similar trends.

% Because of the larger deadlines, the required network and computing resources get reduced with the increased blind distance. However, during the peak hours with the average configuration sometimes for larger blind distance certain jobs miss their deadline and also the response time becomes very large than other lower blind distances. 
Increasing the blind distance requirement generally allows larger response times, thus enabling the deployment of lower edge resources. However, increasing the blind distance too much, i.e., reducing too much the amount edge resources deployed, sometimes have the counter-intuitive effect of leading to higher slow-downs of AVs compared to deploying enough edge resources to provide shorter blind distances. This effect is shown in Figure~\ref{fig:stem}. For example, at 7~am for area A3, the safe speed of AVs decreases from 11~mph to 10~mph when the blind distance is increased from 12~meters to 16~meters.

In general, we also observe a high spatial variation in AV safe speed slowdown. For example, at~8am (another rush hour), it can be seen in Figure~\ref{fig:stem} that for area A3 and A7, even with the average configuration of resources there is no slow down of the AVs with all the considered blind distances. However, as area~A5 has a larger traffic density, the safe speed is reduced to 13~mph, 14~mph, and 17~mph for blind distance of 8~meters, 12~meters, and 16~meters, while the average speed of the regular vehicle is 20~mph. The worst slow down scenario found is at 4~pm, when for area A3, with blind distance 12~meters, the AV's safe speed reduces to 15~mph compared to the regular vehicle speed of 25~mph. This is a 10~mph of slowdown of AVs.

By analyzing the results we can say that using the average configuration of resources at the edge server may effectively handle the traffic demands during all the medium and low traffic hours, which constitute more than two thirds of the day.  During the majority of high traffic hours, we may get a minor slow down of the AV's safe speed. Although there can be higher slowdowns during few rush hours, it can be acceptable given the high cost savings and better resource utilization of the average configuration.

%%% table of the coordinate values goes here

%%Figure having all the scenario result goes here
\begin{figure}[t!]
\vspace{0.1in}
\centering
{\includegraphics[width=\columnwidth]{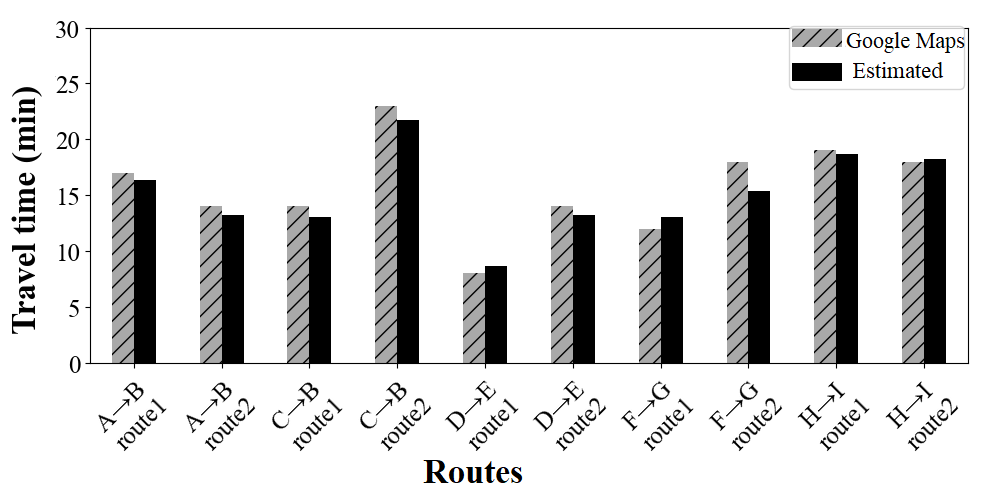}%

\caption{Comparison of the travel time from Google Maps with the estimated time using the data from the dataset for nine different routes.}
\label{fig:travel}}

\vspace{-0.15in}
\end{figure}

\subsection{Travel time: Regular vehicles vs. AVs}
Given the spatio-temporal variability of the safe speed based on time of the day and location for average configurations, here, we investigate whether the faster travel path for regular vehicles provided by modern turn-by-turn navigation algorithms also maps into the faster path for AVs.

\begin{figure*}[!t]
\centering
\subfloat[]{\includegraphics[width=0.33\textwidth,valign=c]{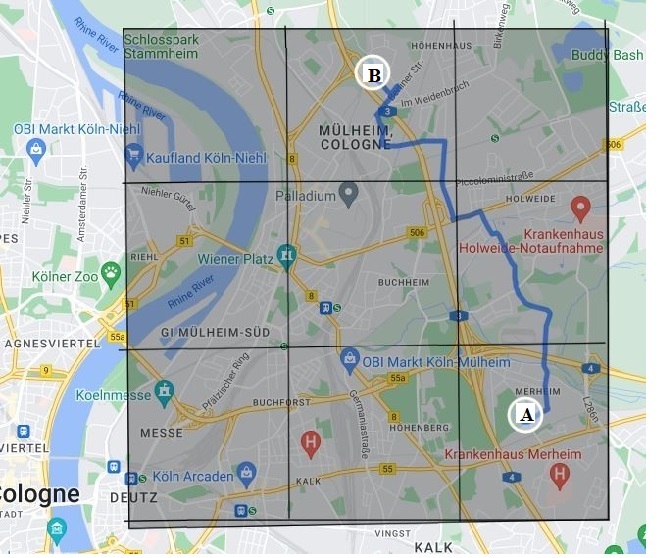}%
\label{fig:sec1route1}
}
\subfloat[ ]{\includegraphics[width=0.33\textwidth,valign=c]{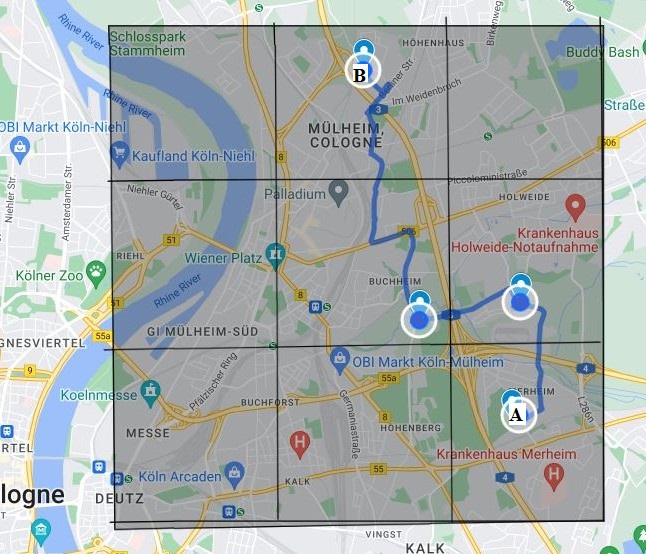}%
\label{fig:sec1route2}
}
\subfloat[ ]{\includegraphics[width=0.33\textwidth,valign=c]{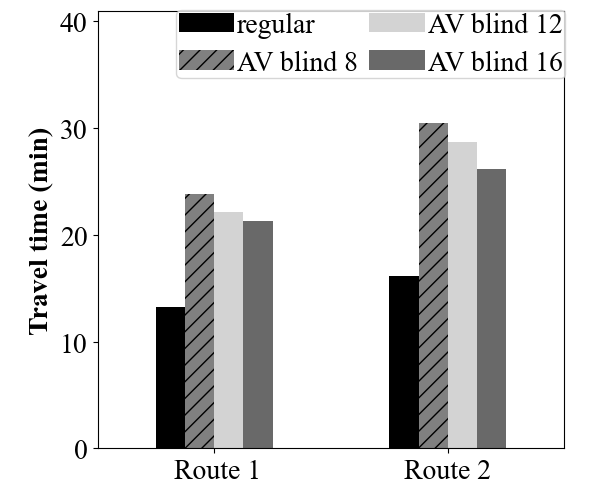}%
\label{fig:sec1}
}
\caption{Scenario I: Going from $A\rightarrow B$ using (a) route 1 given by Google Maps and (b) through an alternative route 2. (c) Comparison of  travel time between the two routes for different blind distances.}
\label{fig: scenario1}
\vspace{-0.17in}
\end{figure*}

\begin{figure*}[!t]
\centering
\subfloat[]{\includegraphics[width=0.33\textwidth,valign=c]{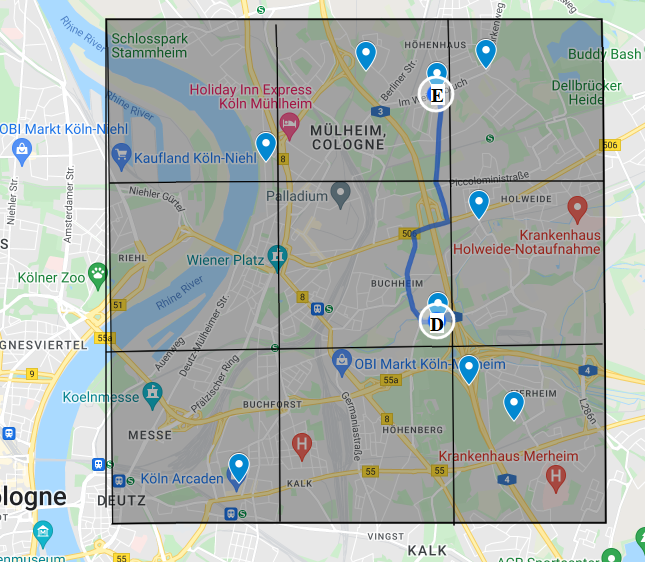}%
\label{fig:sec3route1}
}
\subfloat[ ]{\includegraphics[width=0.33\textwidth,valign=c]{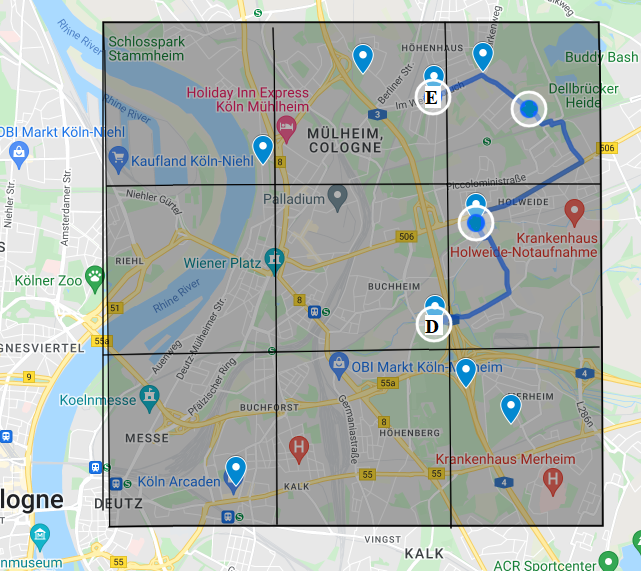}%
\label{fig:sec3route2}
}
\subfloat[ ]{\includegraphics[width=0.33\textwidth,valign=c]{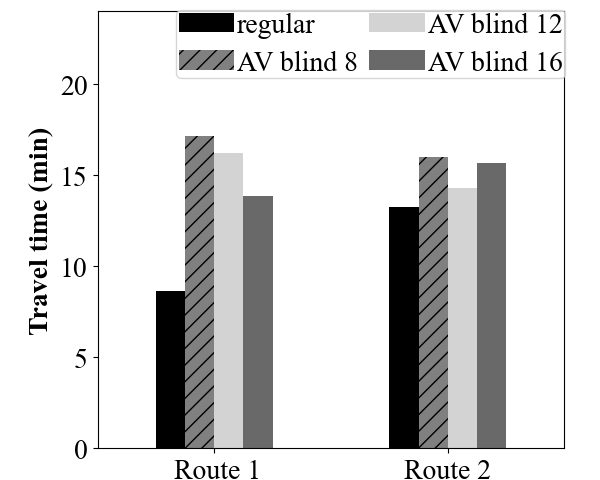}%
\label{fig:sec3}
}
\caption{(a) Scenario II: Going from $D\rightarrow E$ using (a) route 1 given by Google Maps and (b) through an alternative route 2. (c) Comparison of  travel time between the two routes for different blind distances.}
\label{fig: scenario3}
\vspace{-0.17in}
\end{figure*}

\begin{figure*}[!t]
% \vspace{-0.15in}
\centering
\subfloat[]{\includegraphics[width=0.33\textwidth,valign=c]{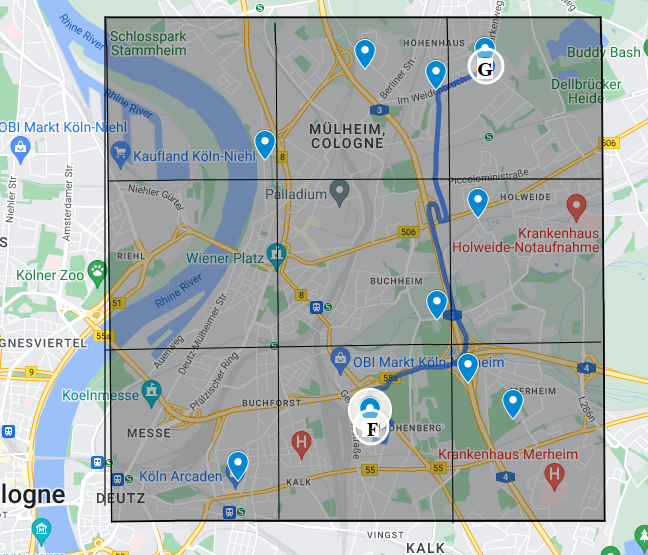}%

\label{fig:sec4route1}
}
\subfloat[ ]{\includegraphics[width=0.33\textwidth,valign=c]{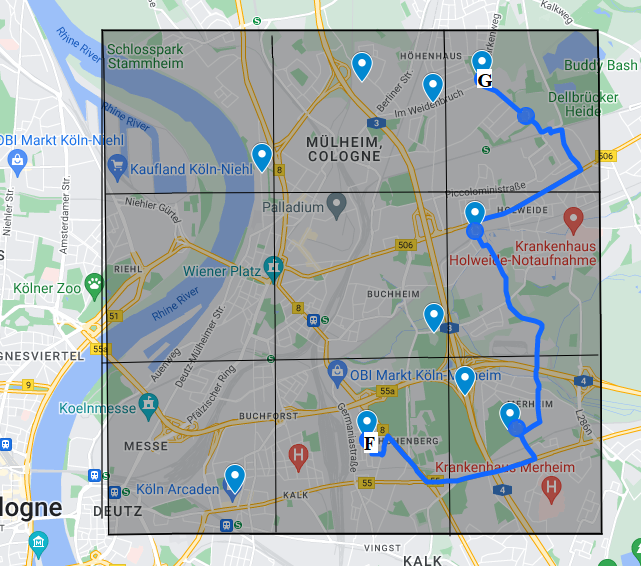}%
\label{fig:sec4route2}
}
\subfloat[ ]{\includegraphics[width=0.33\textwidth,valign=c]{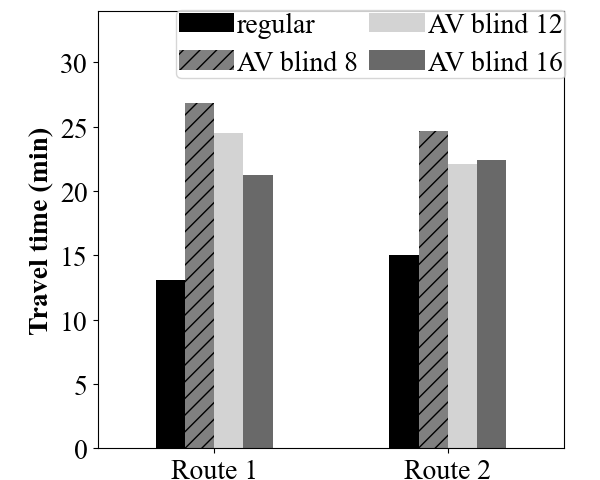}%
\label{fig:sec4}
}
%\vspace{0.1in}
\caption{(a) Scenario III: Going from $F\rightarrow G$ using (a) route 1 given by Google Maps and (b) through an alternative route 2. (c) Comparison of  travel time between the two routes for different blind distances.}
\label{fig: scenario4}
\vspace{-0.17in}
\end{figure*}

In order to conduct this study, we first need to be able to accurately estimate the travel time of edge-assisted AVs on a certain route. To do so, we have created several scenarios where the AVs need to travel from a certain source to a certain destination during rush hour.
%%---- will add the following later--%%
Each scenario has the source and destination located in different areas and two different routes are considered to reach the destination. 
%%---- will add the following later--%%
% One route (Route 1) lets the AV to travel through the heavily loaded areas in terms of traffic and another route takes the AV to go along the light traffic areas.\\
We have compared the travel time shown for each route from Google Maps at 4~pm (one of the rush hours) with our estimated travel time calculated from the average speed from the dataset for different areas at the same day time. We have shown this comparison in Figure~\ref{fig:travel}. Table~\ref{tab:coordinate} contains the location coordinates of the source and destination for the scenarios considered. The results show good estimation accuracy of our method, which validates the results presented in this paper.
% that there is a very low variation in the travel time suggested by Google map and the travel time calculated using the traffic data from the dataset for regular vehicles. So we can consider the results of our further experiments to be nearly correct. 
Among all the scenarios, we show the detailed results for three of them in Figure~\ref{fig: scenario1},~\ref{fig: scenario3}, and~\ref{fig: scenario4}.

%%table having all the coordinates goes here
\begin{table}[t!]
\small
\centering
\caption{Coordinates of the considered locations}
\begin{tabular}{ p{1.5cm}p{3cm}p{1cm} }
 
 \toprule
 \textbf{Location} & \textbf{Coordinates} & \textbf{Area} \\
 \midrule
 
 A  & 50.94571, 7.04523  & A9  \\
 
 B  & 50.98355, 7.01982    & A2  \\
 
 C  & 50.939, 6.99802     & A7  \\
 
 D  & 50.95633, 7.03209    & A5  \\
 
 E  & 50.9812, 7.03192      & A2 \\
 
 F  & 50.94478, 7.02068      & A8  \\
 
 G & 50.98372, 7.04033      & A3  \\
 
 H  & 50.95199, 7.0485     & A9  \\
 
 I  & 50.98346, 6.99847       & A1  \\
\bottomrule

\end{tabular}
\label{tab:coordinate}
\end{table}

%scenario 1

%%%%%%%%%%%
\vspace*{0.1cm}
\noindent\emph{Scenario I.} In Figure~\ref{fig: scenario1}, we consider a scenario where the AVs have to go from location~A, in area~A9, to location~B, in area~A2, at 4~pm on a working day. We consider two different routes to reach the destination. The first route (provided by Google Maps) in Figure~\ref{fig:sec1route1}, requires to go through areas ($A9\rightarrow A6\rightarrow A5\rightarrow A2$) while the second route in Figure~\ref{fig:sec1route2}, takes through the same areas except area~A5 comprises a larger portion than area~A6. The travel distance of route~1 and route~2 are 6.6~km (kilometers) and 7.6~km, respectively. For regular vehicles, the travel time for route~1 is 13~minutes and route 2 is 16~minutes, respectively. Similarly, route~1 is the fastest route also for the AVs. As shown in Figure~\ref{fig:sec1}, the AV can reach the destination 7~minutes, 6~minutes, and 5~minutes earlier considering blind distance of 8~meters, 12~meters, and 16~meters, respectively, via route~1. The reduction in the travel time is due to the increased safe speed of the AVs in area~A6 compared to area~A5 as the traffic density is lower in area~A6.  Thus, in this scenario Google Maps provides the fastest route to both regular vehicles and AVs.

% Thus, we can say that when predicting the time for an AV to reach a certain destination, the edge server configuration along with the other traffic data need to be taken into consideration by any navigation system.

%%%%---figures for three scenarios---%%%

%Fig for scenario 3
\vspace*{0.1cm}

%%for scenario 4
\vspace*{0.1cm}

\noindent\emph{Scenario II.} There can be cases where the fastest route for regular vehicles may not always be the fastest route for edge-assisted AVs. Figure~\ref{fig: scenario3} shows such a case. In this scenario, the source (D) is chosen to be in area~A5 and destination (E) in area~A2. Route~1 ($A5\rightarrow A2$) is the shortest one and it is suggested by Google Maps as the fastest route for regular vehicles, as shown in Figure~\ref{fig:sec3}. However, this route takes the AV to travel through the two heavy traffic areas while route~2 ($A5\rightarrow A6\rightarrow A3\rightarrow A2$) makes the AVs travel through the areas having lower traffic at that time. Although the AV needs to travel a longer distance in route~2,  the increased safe speed allows to lower their travel time compared to that of route~1. It can be seen from the results in Figure~\ref{fig:sec3} that the AV can reach the destination 1~minute and 2~minutes earlier considering blind distance of 8~meters and 12~meters, if it takes the second route. On the other hand, as discussed in the previous section (Figure~\ref{fig:stem}), sometimes the safe speed is reduced with an increased blind distance due to the lower resources necessary on average. As a result, in Figure~\ref{fig:sec3} the travel time increases by 2~minutes while considering blind distance of 16~meters as compared to 12~meters for route~2. This example scenario clearly shows that the travel time for the AVs is dependent not only on the distance but also on the traffic density in the selected route  and the amount of edge resources deployed in different areas throughout the route. Thus, the fastest route shown by the traditional navigation system may not always be the fastest one in case of AVs.

\noindent\emph{Scenario III.} Figure~\ref{fig: scenario4} shows another scenario where the AV has to travel from location~F to location~G. Route~1 (in Figure~\ref{fig:sec4route1}) going through the areas ($A8\rightarrow A9\rightarrow A6\rightarrow A5\rightarrow A2\rightarrow A3$) is shorter and faster for regular vehicles, as suggested by Google Maps where the travel time for the regular vehicles is 13~minutes. However, this route includes the heavy traffic areas A5 and A2. In the case of AVs, with blind distance 8~meters and 12~meters the travel time is approximately 27~minutes and 25~minutes, respectively. If the AV can be rerouted to route~2 shown in Figure~\ref{fig:sec4route2} avoiding the heavy traffic areas, it can be seen from Figure~\ref{fig:sec4} that the travel time can be reduced to 24~minutes and 22~minutes with blind distance 8~meters and 12~meters, respectively. As the length of the second route is greater than the first one, the travel time for regular vehicle becomes longer but in the case of AVs this route takes less time to reach the destination.

Analyzing the results from the different scenarios shown, \emph{we can conclude that while the traditional navigation systems are able to suggest the fastest route for regular vehicles, they are not always efficient in suggesting the fastest routes for the edge-assisted AVs.} 

\section{Future Research Directions}
\label{sec:discuss}
By analyzing all the results shown in Section \ref{sec:results}, we can say that deploying the same configuration of resources at all the edge servers would not be efficient because of different areas having different traffic density at different hours of the day. Furthermore, the transportation system designers may not want to overspend when deploying the edge resources. Thus, deploying the peak configuration of edge resources would not be a reasonable choice as the resources would be used efficiently only during a small fraction of the entire day due to some rush hours and mostly stay idle during other times when the traffic is low. We have seen in the beginning of Section~\ref{sec:results} that the peak configuration requires double the resources compared to the configuration needed to sustain the average demand. Although using the average configuration may result in some higher slowdown in the safe speed of the AVs during the rush hours,  significant reduction of capital cost can be achieved when deploying the resources. Hence, while using the average configuration during the rush hours, efficient measures need to be taken to avoid any unsafe situation by limiting the safe speed of the AVs to guarantee a required blind distance. Thus, \emph{further research should focus on finding an efficient way of provisioning the edge resources so that the capital cost can be reduced while maintaining a desirable system performance. In addition, researchers should also consider how to modify the routing algorithms to take into account the amount of edge resource deployed in different areas and their effect on the expected safe speed for safe and faster navigation of the AVs.}

% We have also seen in Figures~\ref{fig: scenario3} and~\ref{fig: scenario4} that the current navigation systems are not able to always provide the fastest route to reach a certain destination in terms of edge assisted AVs as they do not consider the traffic density and the amount of edge resources deployed can affect the speed of the AVs in certain areas through a particular route. According to that analysis and point of view, we can say that the current turn-by-turn navigation systems are not yet fully ready to be used in terms of edge assisted AVs. 

\section{Conclusion}
\label{sec:conclude}
Motivated by the advantage of edge-assisted autonomous vehicles in providing greater computing power with reduced capital cost, in this paper, we first discussed the implications of deploying different configuration of edge resources in terms of handling peak and average traffic demand. We provided an example system scenario where the AVs offload their heavy computation to the nearby edge servers and described the necessary parameters to consider the offloading execution. Using our algorithms we calculated the required configuration of edge resources to fully handle the hourly demands considering several blind distances. Afterwards, we analyzed the resource requirements for peak and average configurations and found that the peak configuration requires twice as much resources than the average configuration but would only be utilized fully during one third of the entire day. On the other hand, with the average configuration, demand can be met during most of the day while incurring slowdown of AV's safe speed during a fraction of the hours, which could still represent a reasonable option. Finally, we created several scenarios to investigate the time to reach a destination considering the regular vehicles and the edge assisted AVs. With our analysis we found that during the rush hours the AVs can be rerouted to lower their travel time to exploit areas with lower edge-resource utilization. As a result, the traditional turn-by-turn navigation systems in some cases are unable to provide the fastest route to the AVs as they do not take into consideration the amount of edge server resources deployed and the computational delay that may occur in the data processing.


\section*{Acknowledgment}
This research was supported in part by the US National Science Foundation under grants no.\ CCF-2118202 and no.\ CNS-1948365.

\bibliography{main.bib}
\bibliographystyle{IEEEtran}

\newpage
%\section{Biography}
\section*{Biographies}
\vspace{-0.3in}
\begin{IEEEbiography}[{\includegraphics[width=1in,height=1.55in,clip,keepaspectratio]{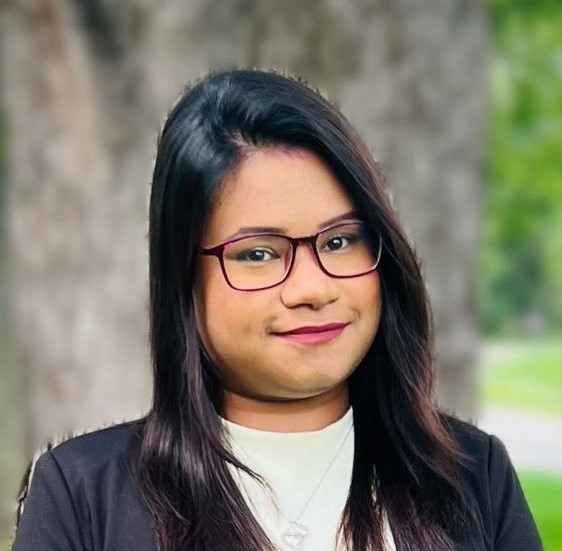}}]
{Syeda Tanjila Atik} is a Ph.D. student in the Department of Computer Science at Wayne State University 
and a student member of the Energy-aware Autonomous Systems Lab (EAS-Lab). She earned her B.Sc. and M.Sc. degrees in
Information Technology from Jahangirnagar University, Bangladesh in 2015 and 2017.  Her research interests are in the area of
edge computing, Internet of Things, autonomous mobile robots, and machine learning.   
\end{IEEEbiography}

\vspace{-0.3in}
\begin{IEEEbiography}[{\includegraphics[width=1in,height=1.25in,clip,keepaspectratio]{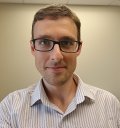}}]
{Marco Brocanelli} (Member, IEEE) is an Assistant Professor in the Department of Computer Science at Wayne State University and the director of the Energy-aware Autonomous Systems Lab (EAS-Lab). He received the B.E. and M.E. degrees in Control Systems from the University of Rome Tor Vergata (Italy). In August 2018, he received the Ph.D. degree in the Electrical and Computer Engineering program of The Ohio State University. His research interests are in the area of cyber-physical systems, energy-aware systems, Internet of Things (IoT), edge computing, embedded and real-time systems. 
\end{IEEEbiography}

\vspace{-0.3in}
\begin{IEEEbiography}[{\includegraphics[width=1in,height=1.25in,clip,keepaspectratio]{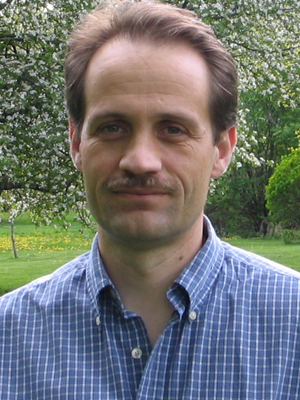}}]
{Daniel Grosu} (Senior Member, IEEE) received the Diploma in engineering (automatic control 
and industrial informatics) from the Technical University of Ia\c{s}i, Romania,
in 1994 and the MSc and PhD degrees in computer science from the University of Texas at
San Antonio in 2002 and 2003, respectively. Currently, he is an associate professor in the
Department of Computer Science, Wayne State University, Detroit.
His research interests include parallel and distributed computing, approximation algorithms, 
and topics at the border of computer science, 
game theory and economics. He has published more than one hundred peer-reviewed papers 
in the above areas. He serves as associate editor and member of the editorial boards of 
\emph{ACM Computing Surveys}, \emph{IEEE Transactions on Parallel and Distributed Systems}, and 
\emph{IEEE Transactions on Cloud Computing}. He is a senior member of the ACM, the IEEE, and 
the IEEE Computer Society.
\end{IEEEbiography}

\vfill

\end{sloppypar}
\end{document}